\documentclass[reprint, amsmath,amssymb, aps]{revtex4-1}
\usepackage{braket}
\usepackage{color}
\usepackage{graphicx}% Include figure files
\usepackage{dcolumn}% Align table columns on decimal point
\usepackage{bm}% bold math
\usepackage{comment}
\newcommand{\eq}{\begin{equation}}
\newcommand{\feq}{\end{equation}}
\newcommand{\virg}[1]{``#1''}

\begin{document}

\preprint{APS/123-QED}

\title{Entangling macroscopic diamonds at room temperature: Bounds on the\\ continuous-spontaneous-localization parameters}

\author{Sebastiano Belli}
\author{Riccarda Bonsignori}%
\author{Giuseppe D'Auria}
\author{Lorenzo Fant}
\author{Mirco Martini}
\author{Simone Peirone}
\affiliation{Department of Physics, University of Trieste, Strada Costiera 11, 34014 Trieste, Italy}
\author{Sandro Donadi}
\author{Angelo Bassi}
\affiliation{Department of Physics, University of Trieste, Strada Costiera 11, 34014 Trieste, Italy}
\affiliation{Istituto
Nazionale di Fisica Nucleare, Trieste Section, Via Valerio 2, 34127 Trieste,
Italy}

\date{\today}% It is always \today, today,
             %  but any date may be explicitly specified

\begin{abstract}
A recent experiment [K. C. Lee {\it et al.}, \textcolor{blue}{Science {\bf 334}, 1253} (\textcolor{blue}{2011})] succeeded in detecting entanglement between two macroscopic specks of diamonds, separated by a macroscopic distance, at room temperature.  This impressive results is a further confirmation of the validity of quantum theory in (at least parts of) the mesoscopic and macroscopic domain, and poses a challenge to collapse models, which predict a violation of the quantum superposition principle, which is the bigger the larger the system. We analyze the experiment in the light of such models. We will show that the bounds placed by experimental data are weaker than those coming from matter-wave interferometry and non-interferometric tests of collapse models.
\end{abstract}

\pacs{Valid PACS appear here}% PACS, the Physics and Astronomy
                             % Classification Scheme.
%\keywords{Suggested keywords}%Use showkeys class option if keyword
                              %display desired
\maketitle
 
%\tableofcontents

\section{\label{sec:level1}Introduction}

The counterintuitive properties of quantum mechanics have always fascinated and puzzled the scientific community. This is even more true now that rapid technological developments allow or promises to test quantum physics in regimes and conditions, which were unaccessible only a few years ago \cite{aa,romero}.  At stake there is not only a deeper understanding of nature, but also the hope to set the ground for novel quantum technologies.

The quantum-to-classical transition certainly is the most problematic aspect of quantum theory. Why is our direct physical experience that of a classical world, if we are made of atoms and molecules, which obey the laws of quantum physics? Why do we not see superpositions and entanglement in play in everyday life? Can these be pushed to the macroscopic domain, at least in controlled environments such as laboratory experiments? 

A recent experiment \cite{lee} made a significant step forward in this direction. Two millimeter-size diamonds, distant some 15 cm one from the other, were entangled, and the entanglement was detected, via optical techniques, at room temperature. This means that objects directly visible  by the human eye, in standard environmental conditions, and separated by a macroscopic distance, show a quantum behavior. 

This remarkable result, besides being interesting for our understanding of nature, poses a challenge to collapse models \cite{grw,pearle,gpr,diosi,gatarek,chru,bass2}. These models predict an evolution for the state vector fundamentally different from standard Quantum Mechanics: every system interacts non linearly with a classical noise, which induces the collapse of the wave function in space. This violation of quantum linearity depends on the size of the system: it is negligible for microscopic systems, and increases with the number of constituents to the point that macroscopic objects are always well-localized in space. Somewhere in the mesoscopic domain, the breakdown of the quantum superposition principle starts becoming significant, explaining the quantum-to-classical transition.

The question we address here, is what the experimental result has to say regarding these models. The lifetime of the entangled state is very short ($\sim 10^{-12}$s), but on the other hand the masses ($\sim 10^{16}$amu) and distances ($\sim 10$cm) involved are very large, truly macroscopic, and could in principle compensate the very short time. On the other hand, it is not immediately clear that collapse models should enter into play here at all. There is no center-of-mass motion involved, which is the typical way to enhance the collapse effect and test it experimentally. On the other hand, internal vibrations imply mass motion, to which these models are sensitive. 

We will clarify all these issues. In our analysis, we will consider the mass-proportional Continuous Spontaneous Localization (CSL) model~\cite{gpr}, which is the reference model in the literature for the comparison with experimental data. 
People refer to this model for historical and practical reasons. It was the first model including the description of identical particles; moreover, the collapse was formulated in terms of a continuous diffusion process, as opposed to the original Ghirardi-Rimini-Weber (GRW)~\cite{grw} model, the first consistent collapse model, where the collapse is described by a discrete jump process. Continuous processes are somehow easier to work with.
 We will compute the predictions that CSL makes for the experiment under consideration, and which bounds the data place on the CSL parameters.

%%%%%%%%%%%%%%%%%%%%%%%%%%%%%%%%%%%%%%%%%%%%%%%%%%%%%%%%%%%%%%%%%%%%%%%%%%%%%%%%%%%
\section{Experimental Setup}
\begin{figure}[t]
\centering
\includegraphics[scale=0.35]{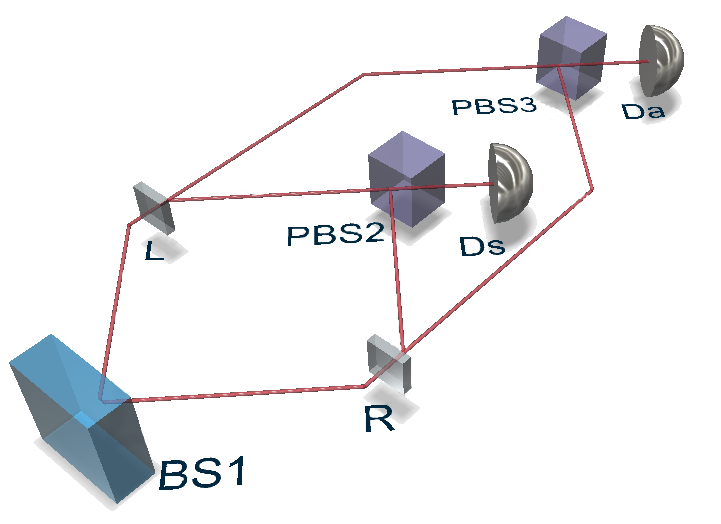}
\caption{Experimental setup, schematic representation. A pump pulse, split by the beamsplitter $BS1$, is sent to the diamonds. The two separate Stokes modes produced through Raman transitions and exiting each diamond, are  combined  by the polarization beamsplitter $PBS2$ and reach the detector $D_s$. A probe pulse sent after a short time interval is sent through the diamonds to produce anti-Stokes photons, which are then combined on $PBS3$ and sent to the detectors $D_{a}$.
\label{setup}}
\end{figure}

%%%%%%%%%%%%%%%%%%%%%%%%%%%%%%%%%%%%%%%%%%%%%%%%%%%%%%%%%%%%%%%%%%%%%%%%%%%%%%%%%%%
The experiment is described in \cite{lee} and reviewed in Appendix~\ref{app1}. In 
this section we give a concise presentation of the experimental setup, which is summarized in Fig.~\ref{setup}. The core consists of two 0.25mm thick diamond plates (3mm $\times$ 3mm in size), which are spatially separated by a distance of 15cm. Entanglement is created among their phononic states, using pump-and-probe ultrashort optical pulses with a bandwidth of $\sim$~7THz. 

The initial pump pulse is split in two by a 50:50 beamsplitter (BS1), and the two parts are sent each to one diamond. When the pulse is absorbed, a phonon is created inside the diamond, and a Stokes photon with wavelength $\lambda_s$= 900nm is emitted.  The optical phonon mode is a bulk vibration consisting of two counter-oscillating sub-lattices of $10^{16}$ atoms in a volume of side $L \sim 10^{-5}$m, with a carrier frequency of 40THz. 

The Stokes photons emitted by each diamond are recombined by a polarizing beam splitter (PBS2) and interfere with each other through an half-wave plate and polarizer with relative phase shift $\phi_s$. Then they are detected by a single photon counter. 

To verify that the detection of the Stokes photons is evidence of entanglement between the two diamonds, strong probe pulses are directed to the crystals at a time $t_a = t_s + T$ with T = 350fs, before each Stokes photon has reached the detector, where $t_s$ is the time when the Stokes photon is created.

The interaction of the probe pulse with the diamond induces the transition of the phononic state to the ground state, with the emission of an anti-Stokes photon with wavelength of $\lambda_a$= 735nm. Due to their different frequencies, the optical paths of the Stokes and anti-Stokes photons can be separated into two different spatial modes by means of a long pass filter. 

The anti-Stokes photons are then combined through a beamsplitter, interfere with each other through an half wave plate and polarizer with a relative phase shift of $\phi_a$ and finally detected. The role of the probe pulse is to coherently transfer the entangled phonon state into the anti-Stokes mode for entanglement verification. 

%%%%%%%%%%%%%%%%%%%%%%%%%%%%%%%%%%%%%%%%%%%%%%%%%%%%%%%%%%%%%%%%%%%%%%%%%%%%%%%%%%%
\section{CSL prediction and bounds on the collapse parameters}
\begin{figure}[b]
\centering
\includegraphics[width=0.45\textwidth]{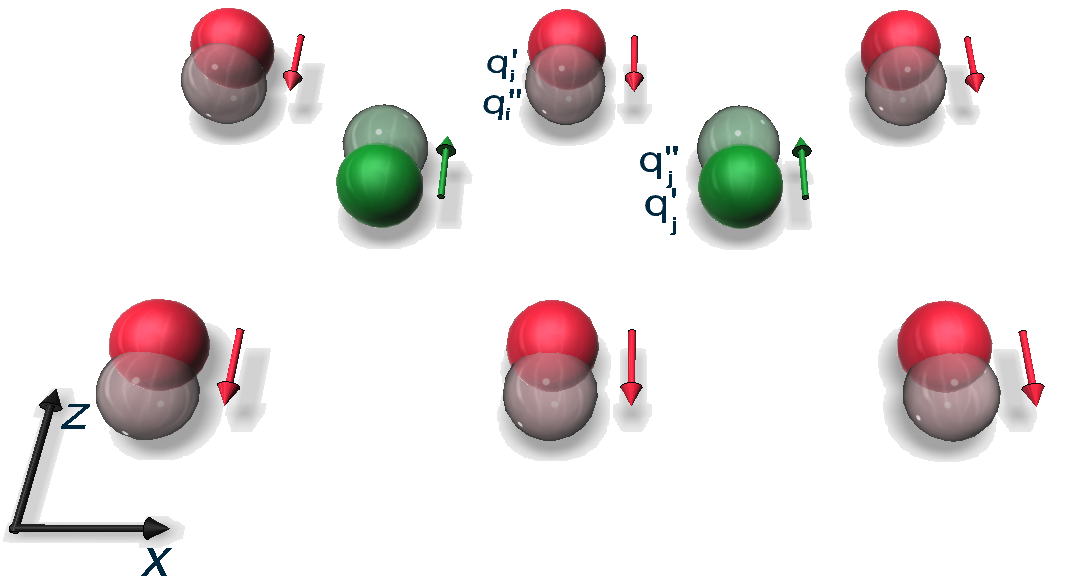}\label{Fig:3}
\caption{Schematic representation of the displacements of carbon atoms in one diamond. The two different colors represent the two different counter-oscillating sublattices. Full colors denote  the two sublattices at rest, while faded ones represent the two sublattices at the maximum relative distance. %There are two counter-oscillating sublattices and  the circles and the crosses represent respectively one sublattice and the other one. 
 \label{sublatt}}
\end{figure}

%%%%%%%%%%%%%%%%%%%%%%%%%%%%%%%%%%%%%%%%%%%%%%%%%%%%%%%%%%%%%%%%%%%%%%%%%%%%%%%%%%%
Apart from all (important) details, the essence of the experiment is that it creates the superposition of diffe\-rent matter distributions inside the two crystals, as shown in Fig.~\ref{sublatt}.
Differently from other interferometric proposal for testing CSL effects \cite{grw,pearle,gpr,diosi,gatarek,chru,bass2,bass3,bass4}, which all rely on creating the center-of-mass superpositions of massive systems, in this case the center of mass of each of the two crystals remains always well localized. However, within each diamond, two sublattices are placed in the superposition of either being at rest (no Stokes photon produced) or of oscillating (Stokes photon produced). Therefore we have a superposition of different matter distributions in space. Since the CSL dynamics is constructed precisely to destroy superpositions of matter in space, and the effect scales with the mass of the system, we expect the dynamics of the experiment to be sensitive to CSL effects.
We now compute these effects.

The mass-proportional CSL master equation for the density matrix $\rho_t$, in the position representation, is \cite{bass2}:
\eq
\label{eq:4}
\begin{aligned}
 \frac{d}{d t} \rho_t \, =& - \frac{i}{\hbar}\,  \left[ H,\rho_t \right ] \,
 \\
 &- \frac{ \lambda}{2 r_C^3 \, \pi^{3/2} \, m_0^2}  \int  d {\bf x} \, \left [ M({\bf x}),\left[M({\bf x}) ,\rho_t \right]  \right] \,,
\end{aligned}
\feq
where $m_0 = 1$amu and: 
\eq
\label{eq:5}
M({\bf x})= \sum_n \, m_n e^{-({\bf x} - \hat {\bf x}_n)^2/2 r_C^2}\,,
\feq
where $\hat {\bf x}_n$ denotes the position operator of the $n$-th particle, and the sum runs over all particles. The  model is characterized by two phenomenological constants, a collapse rate $\lambda$ and a characteristic length $r_C$, which measure respectively the intensity and the spatial resolution spontaneous collapse.
The standard values suggested for CSL parameters are $\lambda=  10^{-17}$s$^{-1}$ and $r_C= 10^{-7}$m~\cite{bass2}. 
A  value about 8 orders of magnitude stronger for the collapse rate has been suggested by Adler \cite{adler}, motivated by the requirement of making the wave function collapse effective at the level of  latent image formation in photographic process. 

In Eq.~\eqref{eq:5}, the sum is taken over all the particles, but being $m_0$ the mass of the nucleon, the terms associated to the electrons contain a multiplicative factor $ ( m_e  / m_0)^2~\sim~10^{-6}$ and therefore they can be neglected.

We  label with \virg{$L$} and \virg{$R$} respectively the diamond at left and right, and the two counter oscillating sublattices with $1$ and $2$.  $M(\textbf{x})$ can be rewritten in terms of sum over the set of atoms belonging to each  sublattice:
\eq
\label{massa.1}
M(\textbf{x})= \sum_{\alpha} \sum_{n\in A_{\alpha}} m_n e^{-(\textbf{x} - \hat{\textbf{x}}_n)^2/2r_C^2}\,,
\feq
where $\alpha = L_1,L_2,R_1,R_2$ is the index of sublattice and $A_{\alpha}$ is a set of atoms belonging to the same sublattice.

The following expression holds:
\eq
\hat{\textbf{x}}_n=  \textbf{x}^{(0)}_{n\,,\alpha} + \hat{\textbf{x}}_{\alpha} + \Delta \hat{\textbf{x}}_n\,,
\feq
where $\textbf{x}^{(0)}_{n\,,\alpha}$ is the classical rest position of particle $n$ in sublattice $\alpha$, while $\hat{\textbf{x}}_{\alpha}$  marks the oscillation of the sublattice (here we are assuming that the sublattice moves rigidly when excited), and $\Delta \hat{\textbf{x}}_n$ denotes the displacement operator of the particle due to the quantum motion around the equilibrium position. Since these oscillations are very small compared to the inter-particle distance ($\sim$~2 orders of magnitude smaller), we can assume the rigid body approximation and neglect the last term. Therefore we have:
\eq \label{eq:em}
M({\bf x}) = \sum_{\alpha} \int d \textbf{r}\, \mu_{\alpha}(\textbf{r}) e^{-(\textbf{x} -  \hat{\textbf{x}}_{\alpha}  -\textbf{r})^2/2r_C^2}\,,
\feq 
where we have introduced the mass density distribution of the sublattice:
$
\mu_{\alpha}(\textbf{r})= \sum_{n \in A_{\alpha}} m_n \delta^{(3)} (\textbf{r}- \textbf{x}^{(0)}_{n\,,\alpha})\,.
$

Let us define  $g(\hat{\textbf{x}}_{\alpha}) = \exp[- (\textbf x - \hat{\textbf{x}}_{\alpha} - \textbf r)^2 /2 r_C^2]$. Since the sublattice vibrates a little around the equilibrium,  $\hat{\textbf{x}}_{\alpha}$ is small compared to the other variables, justifying a Taylor expansion around $\hat{\textbf{x}}_{\alpha} = 0$:
\eq \label{21}
g(\hat{\textbf{x}}_{\alpha}) \approx g(0) + \nabla \left. g(\hat{\textbf{x}}_{\alpha}) \right|_{\hat{\textbf{x}}_{\alpha}=0} \cdot \hat{\textbf{x}}_{\alpha}\,.
\feq
Using the Taylor expansion we can rewrite $M(\textbf{x})~=~M_0(\textbf{x}) + \hat{M}_1(\textbf{x})$, where $M_0$ contains no operator and $\hat{M}_1$ is linear in $\hat{\textbf{x}}_{\alpha}$.
Then the second term of Eq.~\eqref{eq:4} becomes:
\eq
\label{eq12}
\begin{aligned}
-\frac{\lambda}{2r_{C}^{3}\pi^{3/2}m_{0}^{2}}&\sum_{\alpha,\alpha'}\int d\mathbf{r}_{1}\int d\mathbf{r}_{2}\,\mu_{\alpha}(\mathbf{r}_{1})\,\mu_{\alpha'}(\mathbf{r}_{2})
\\
&\times I_{ij}(\mathbf{r}_{1},\,\mathbf{r}_{2})\left[(\hat{\mathbf{x}}_{\alpha})_{i},\left[(\hat{\mathbf{x}}_{\alpha'})_{j},\rho_{t}\right]\right],
\end{aligned}
\feq
with:
\eq 
\label{Sandrino}
I_{ij}(\mathbf{r}_{1},\!\mathbf{r}_{2}) \!= \!\frac{1}{r_{C}^{4}} \!\!\int \!\! d\mathbf{x}e^{-\left[(\mathbf{x}-\mathbf{r}_{1})^{2}\!+\!(\mathbf{x}-\mathbf{r}_{2})^{2}\right]/2r_{C}^{2}}(\mathbf{x}-\mathbf{r}_{1})_{i}(\mathbf{x}-\mathbf{r}_{2})_{j} .
\feq

Since  $\mu_{\alpha}$ and $\mu_{\alpha'}$ refer to different spatial regions, the
contributions of the terms with $\alpha\neq\alpha'$ in Eq.~\eqref{eq12} are negligible. 
Moreover, since the phonon sets the sublattices in motion only along the $z$-direction ($i=j=3$), the only relevant contribution in Eq.~\eqref{eq12} comes from
$\left[(\hat{\mathbf{x}}_{\alpha})_{3},\left[(\hat{\mathbf{x}}_{\alpha})_{3},\rho_{t}\right]\right]:= \left[z_{\alpha},\left[z_{\alpha},\rho_{t}\right]\right].$
After  simple calculations, which are reported in Appendix~\ref{app2}, we can rewrite the expression in Eq.~\eqref{eq12} as:
\eq
- \sum_{\alpha} \eta_{\alpha} [z_{\alpha},  [z_{\alpha}, \rho_t ] ]\,,
\feq
with:
\begin{eqnarray}
\label{eta}
\eta_{\alpha} & = &  \frac{\lambda}{4 r_C^4 m_0^2 } \int d \textbf r_1 d\textbf r_2 \mu_{\alpha}(\textbf r_1)\mu_{\alpha}(\textbf r_2)\, e^{ -(\textbf r_1 - \textbf r_2)^2/4 r_C^2} \nonumber
\\
& & \times\left[ r_C^2 - \frac{\left(z_1 - z_2 \right)^2}{2} \right]\,.
\end{eqnarray}

As  previously pointed out, the optical phonon modes are characterized by the counter oscillations of the sublattices $1$ and $2$ inside each diamond. This means that, at every time:
\eq
\begin{aligned}
&\hat {z}^L_1=- \hat {z}^L_2 =\hat {q}^L\, , \qquad \hat {z}^R_1=- \hat {z}^R_2 =\hat {q}^R.
\end{aligned}
\feq
Taking also into account that the mass distribution is the same for each sublattice  ($\eta_{\alpha} = \eta, \, \forall {\alpha}$), we can rewrite Eq.~\eqref{eq:4} as follows:
\eq
\label{29}
 \frac{d}{dt}  \rho_t  = - \frac{i}{\hbar}\,  [ H,\rho_t ] - 2 \eta [\hat {q}^L, [\hat{q}^L, \rho_t]]  -  2 \eta [\hat {q}^R, [\hat{q}^R, \rho_t ]] ,
\feq
which has the well-known form first derived by Joos and Zeh in \cite{jo-ze}. The first term on the right encodes the standard quantum evolution, as previously described. The remaining two terms contain the CSL effect.

In order to solve Eq.~\eqref{29}, we first express $\hat {q}^{L,R}$ in terms of the annihilation and creation operators, both for the left $L$ and right $R$ case:
\eq
\hat {q}^L  =\sqrt{\frac{\hbar}{\omega m^* }} \frac{\hat a_L + \hat a_L^{\dagger}}{\sqrt{2}}\,, \qquad \hat {q}^R =\sqrt{\frac{\hbar}{\omega m^* }} \frac{\hat a_R + \hat a_R^{\dagger}}{\sqrt{2}}  \,,
\feq
where $m^*=6\, m_0$ is the reduced mass of the unit cell~\cite{prive}.
We restrict the analysis to the 4D subspace, tensor product of the two  2D ``left'' and ``right'' subspaces generated by the vacuum and the single-phonon states.
In other words, we compute the matrix elements of Eq.~\eqref{29} restricting the analysis to the states:
\eq
\begin{aligned}
\label{eq21}
& | 0_L, \, 0_R \rangle = | 0 \rangle \,, \qquad \hspace*{4mm}| 1_L, \, 0_R \rangle = \hat a_L^{\dagger}| 0 \rangle \,, \\
& | 0_L, \, 1_R \rangle = \hat a_R^{\dagger} | 0 \rangle \,, \qquad |1_L, \, 1_R \rangle = \hat a_L^{\dagger} \, \hat a_R^{\dagger} | 0 \rangle  \,.
\end{aligned}
\feq
Eq.~\eqref{29} then becomes:
\eq \label{33}
\frac{d}{dt}  \rho_t \, = - i \, \omega\,  \left[ H,\rho_t \right ] \, - \, \frac{1}{2} \, \Lambda\, \Gamma[\rho_t]\,,
\feq
\noindent
where $\Lambda = 4 \eta \hbar / \omega m^*$ and the matrix elements of the operators $[H, \rho]$ and $\Gamma[\rho]$, computed in the basis defined in Eq.~\eqref{eq21}, are:

\begin{widetext}

\eq 
\label{35.}
[H, \rho]=  \left( \begin{matrix} 
 0 ; & -\rho_{12} ; & -\rho_{13} ; & -2 \, \rho_{14} \\
\rho_{21} ; & 0 ; & 0 ; & - \rho_{24} \\
\rho_{31} ; & 0 ; & 0 ; & - \rho_{34} \\
2 \, \rho_{41} ; & \rho_{42} ; & \rho_{43} ; & 0 \\
 \end{matrix}\right)
\feq 
\eq 
\label{34.}
\Gamma[\rho]= \left( \begin{matrix} 2 \rho_{11} - \rho_{22} - \rho_{33}; & 2 \rho_{12}-\rho_{21}-\rho_{34} ; & 2 \rho_{13} - \rho_{24} -\rho_{31} ; & 2 \rho_{14} -\rho_{23}-\rho_{32} \\ 2 \rho_{21} - \rho_{12} -\rho_{43} ; & 2 \rho_{22}-\rho_{11}-\rho_{44} ; & 2 \rho_{23}-\rho_{41}-\rho_{14} ; & 2 \rho_{24}-\rho_{13}-\rho_{42}  \\ 2 \rho_{31}-\rho_{42}-\rho_{13} ; & 2 \rho_{32}-\rho_{41}-\rho_{14} ; & 2 \rho_{33}-\rho_{44}-\rho_{11} ; & 2 \rho_{34} -\rho_{43} -\rho_{12} \\ 2 \rho_{41}-\rho_{32}-\rho_{23} ; & 2\rho_{42}-\rho_{31}-\rho_{24} ; & 2\rho_{43}-\rho_{34}-\rho_{21} ; & 2 \rho_{44}-\rho_{33} -\rho_{22} \end{matrix}\right)
\feq 

\end{widetext}

In accordance with the description of the previous section, we take as initial state the superposition of the states $| 1_L, \, 0_R \rangle$ and $| 0_L, \, 1_R \rangle$, which in the density matrix formalism gives the following initial condition:
\eq 
\rho_0= \frac{1}{2}\left( \begin{matrix}
0 & 0 & 0 & 0\\
0 & 1 & 1 & 0\\
0 & 1 & 1 & 0\\
0 & 0 & 0 & 0\\
\end{matrix} \right).
\feq
The presence of the off diagonal elements are a measure of quantum superposition and, in this case, of entanglement.
The time evolution of the matrix elements, according to Eq.~\eqref{33}, is:
\eq 
\begin{aligned}
\nonumber
\rho_{11}=&\rho_{44} = \frac{1}{4} \, \left( 1 - e^ { - 2 \Lambda t} \right),\\
\rho_{22}=&\rho_{33} =  \frac{1}{4} \, \left( 1+ e^ { - 2 \Lambda t} \right),\\
\rho_{14}=& \frac{\Lambda}{2} e^{-\Lambda t}  \,  \bigg(\frac{2 i \omega \left[\cosh(\Omega \, t)-1\right]}{\Omega^2} +\frac{\sinh(\Omega \, t)}{\Omega} \bigg)\\
\rho_{41}=& \frac{\Lambda}{2} e^{-\Lambda t} \, \bigg(-\frac{2 i \omega \left[\cosh( \Omega \, t)-1\right]}{\Omega^2} +\frac{\sinh( \Omega \, t)}{\Omega} \bigg)\\
\rho_{23}=&\rho_{32}=\frac{1}{2} e^{-\Lambda t} \, \frac{\Lambda^2 \cosh( \Omega \, t) - 4 \omega^2}{\Omega^2}\\
\end{aligned}
\feq
where $\Omega=\sqrt{\Lambda^2 -4 \omega^2}$ and the other terms are zero. 

It is interesting to observe that in the long-time limit the density matrix approaches:
\eq 
\rho_{\infty} =\frac{1}{4}\left( \begin{matrix}
1 & 0 & 0 & 0\\
0 & 1 & 0 & 0\\
0 & 0 & 1 & 0\\
0 & 0 & 0 & 1\\
\end{matrix} \right).
\feq
As we can see, the off-diagonal elements are suppressed, as an effect of the collapse mechanism embodied in the second term of Eq.~\eqref{eq:4}. Moreover, all states are equally populated, as an effect of the noise which, on top of killing superpositions, also induces a Brownian-type of motion inside the crystals, where phononic states are constantly created and destroyed. 

The next step now is to evaluate $\eta$. To do so we choose to follow the approach used in \cite{nhh}, with some obvious differences.  Starting from Eq.~\eqref{massa.1}, we write the function $M(x)$ in Fourier space as:
\eq
\begin{aligned}
\nonumber
M(\textbf{x})=& \sum_{\alpha} \sum_{n\in A_{\alpha}} m_n e^{-(\textbf{x} - \hat{\textbf{x}}_{\alpha} - \hat{\textbf{x}}_n)^2/2r_C^2} \\
=&\frac{r_C^3}{(2 \pi )^{3/2}} \sum_{\alpha}\int\, d {\bf k} \, e^{- \frac{r_C^2 k^2 }{2}} \, e^{ i \, {\bf k} \cdot ({\bf x} - {\bf \hat x}_{\alpha})} \tilde \mu ({\bf k})
\end{aligned}
\feq
where  $\tilde{\mu }({\bf k})$ is the Fourier transform of the density $\mu ({\bf r})$, which is the same for all sublattices. From~Eq.~\eqref{eta} we obtain, after tedious but straightforward calculations, the expression of $\eta$ as a Fourier integral:
\eq
\eta= \frac{\lambda \, r_C^3}{2 \, \pi ^{3/2} m_0^2} \, \int \, d {\bf k} \, k_z ^2 \left| \tilde \mu({\bf k}) \right| ^2 \, e^{- r_C^2 k^2}.\\
\feq

In the  experiment here considered, each phonon has roughly a cylindrical shape with a radius $R=~3.6 \mu$m and length equal to the width of the diamonds, $d=~0.25$mm \cite{prive}. Since the  density of atoms for diamonds is $n=~176.2 \times 10^{27}$m$^{-3}$, the total number of atoms contributing to the phonon is $N~\sim~2~\times~n~\times~( \pi \, R^2 \, d)=~3.6~\times 10^{15}$ (the factor $2$ arises because there are two diamonds).
For a cylinder with homogeneous mass density:
\begin{align}
 \mu({\bf r}) = &
\begin{cases}
   \frac{m}{\pi \, R^2 \, d} \qquad {\bf r} \in V, \\
   0 \qquad \qquad \text{otherwise,}
\end{cases}
\end{align} 
and its Fourier transform  is  \cite{nhh}:
\eq
\tilde{\mu}_{\text{\tiny cyl}}(k_z, {\bf k_{\perp}}) = \frac{2\, m}{ k_{\perp} \, R} \, J_1 (k_{\perp}\, R)\, \text{sinc} \left( \frac{k_z \, d}{2} \right),
\feq
where ${\bf k_{\perp}} = (k_x, k_y)$, $J_1$ denotes the Bessel Function and $m= 12 \times N \times m_0 $ is the total mass (the factor 12 comes from the carbon atom mass number $A=12$). Finally we arrive at  the expression:
\eq \label{etafinal}
\eta = \lambda \frac{N^2}{d^2} \Gamma_{\perp}\left(\frac{R}{\sqrt{2}\, r_C}\right)\, \left[ 1 - e^{- \frac{d^2}{4\, r_C^2}} \right],
\feq
where:
\eq 
\Gamma_{\perp} (x)= \frac{2}{x^2} \left[ 1- e^{-x^2} \, \left( I_0(x^2) +I_1(x^2) \right) \right],
\feq
with $I_0$ and $I_1$ the modified Bessel functions.

\begin{comment}
Now, as clearly presented in \cite{nhh}, we consider the mass density distribution in Fourier space of an homogeneous sphere of radius $L$:
\eq
\tilde{\mu({\bf k})} = 3 m \frac{\sin(k L) - k L \cos(k L)}{(k L)^3}
\feq
where $m= N \, m_0$ is the total mass of the system. Finally we get then the expression:
\eq 
\nonumber
\eta = \lambda \frac{6   r_C^3}{  \pi ^{1/2}}\left(\hspace{-1mm}\frac{m}{m_0}\hspace{-1mm}\right)^2  \hspace{-1mm} \int_0^{\infty} \hspace{-2mm} d { k}  k^4 \left| \frac{\sin(k L) - k L \cos(k L)}{(k L)^3} \right| ^2 \hspace{-1mm} e^{- r_C^2 k^2}\hspace{-1mm}.
\feq
Doing calculation, we get:
\eq 
\nonumber
\eta = \lambda \frac{3 \, r_c^4}{L^6} \left(\frac{m}{ m_0}\right)^2  \left[ e^{-\frac{L^2}{r_c^2}} -1+\frac{L^2}{2r_c^2}\left(e^{-\frac{L^2}{r_c^2}}+1\right)\right].
\feq
\end{comment}
Taking  the reference value $r_C = 10^{-7}$m~\cite{bass2}, and by plugging in all the numbers, we get:
\eq \label{eq:asd}
\eta \sim 6 \times 10^{35} \,\lambda \, \text{m}^{-2} 
\feq

\begin{figure}[t]
\centering
\includegraphics[scale=0.31]{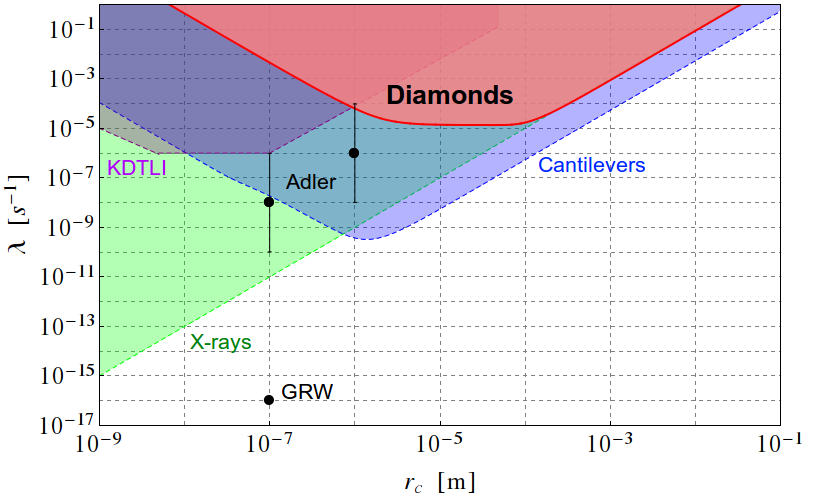}\label{Fig:4}
\caption{Exclusion plot in the $\lambda$-$r_C$ parameter space. The red curve (``Diamonds") shows the bound here computed from the experiment under consideration; the region marked in red is excluded with a high confidence level. For comparison we have included the upper bounds from X-ray experiment \cite{cc} (green), from nanocantilever experiments \cite{bb} (blue) and  from KDTLI experiment \cite{aa} (purple). For reference, we have also included the GRW \cite{grw} values ($\lambda=10^{-16}$s$^{-1}$, $r_C=10^{-7}$m) and the values proposed by Adler \cite{adler}: ($\lambda=10^{-8}$s$^{-1}$, $r_C=~10^{-7}$m) and ($\lambda=10^{-6}$s$^{-1}$, $r_C=10^{-6}$m).}\label{excl}
\end{figure}

To extract the order of magnitude of the CSL effect during the measurement, let us take the solutions of Eq.~\eqref{29} and neglect the contribution of $H$.
Then,  the density matrix in the position representation evolves as:
\eq 
\rho_t= \rho_0 \, e^{-4 \, \eta \,\left( \Delta z \right)^2  T}\,,
\feq
where $\Delta z = \sqrt{\frac{\hbar}{m^* \omega}} = 1.6 \times 10^{-11} $m is the maximum displacement due to the oscillation of the phonon in one diamond~\cite{prive} and $T=350 $fs. Since the experiment shows full quantum interference which implies no sign of collapse effects, the following relation must be true:
\eq 
4 \, \eta \,\left( \Delta z\right)^2 \, T\ll 1,
\feq 
from which we identify the exclusion zone reported in Fig.~\ref{excl} (red region). 

The shape of the exclusion zone can be understood by studying the dependence of $\eta$ on $\lambda$ and $r_C$. From Eq.~\eqref{etafinal} we see that $\eta$ always depends linearly on $\lambda$, while the dependence on $r_C$ changes for different values of $r_C$. More precisely (keeping in mind that $R=3.6\times~10^{-6}$m and $d=2.5\times10^{-4}$m) it can be shown that when $r_{C}\leq~10^{-7}\textrm{m}$ then $\eta\sim\lambda r_{C}^{2}$ (slope = $-2$ for the red line), when $r_{C}\simeq10^{-5}\textrm{m}$ (slope = 0) then $\eta\sim\lambda$ and when $r_{C}\geq10^{-3}\textrm{m}$ then $\eta\sim\lambda/r_{C}^{2}$ (slope = 2).    

%%%%%%%%%%%%%%%%%%%%%%%%%%%%%%%%%%%%%%%%%%%%%%%%%%%%%%%%%%%%%%%%%%%%%%%%%%%%%%%%%%%
\section{Comments and conclusions} 
%%%%%%%%%%%%%%%%%%%%%%%%%%%%%%%%%%%%%%%%%%%%%%%%%%%%%%%%%%%%%%%%%%%%%%%%%%%%%%%%%%%
The bound here derived is weak in the sense that neither Adler's value for $\lambda$, nor GRW's (black points in Fig.~\ref{excl}), are tested by the experiment. The reason is easy to understand.  First of all, entanglement (over a distance of 15cm) here does not play a role. The reason is that the real superposition of matter is within the two crystals (the two counter-oscillating sub-lattices), not between the two diamonds. CSL is insensitive to entanglement, which does not carry spatial superposition of matter. The second reason is that  the superposition times and distances are too small, and are not fully compensated by the very large number of atoms involved in the superposition.

In spite of this, our bound competes with that coming from KDTLI~\cite{aa}, where the superposition times and distances are larger, but the number of atoms involved is much smaller. These two are, so far, the only direct, i.e. interferometric, tests of the quantum superposition principle, in the sense that they both create and test a quantum superposition of the center-of-mass position of a relatively massive system. The latter is stronger for smaller values of $r_C$, ours for larger values.

Both bounds, however, are weaker than those coming from non-interferometric tests: violation of the equipartition theorem with cantilevers ~\cite{bb}, and  X-ray measurements~\cite{cc}. At present, non-interferometric tests proved the strongest upper bound for the CSL parameters; what is not clear is whether these bounds are robust against  changes in the collapse mechanism, while those associated to interferometric tests are~\cite{mt}.

%%%%%%%%%%%%%%%%%%%%%%%%%%%%%%%%%%%%%%%%%%%%%%%%%%%%%%%%%%%%%%%%%%%%%%%%%%%%%%%%%%%
\section*{Acknowledgements}
%%%%%%%%%%%%%%%%%%%%%%%%%%%%%%%%%%%%%%%%%%%%%%%%%%%%%%%%%%%%%%%%%%%%%%%%%%%%%%%%%%%

The authors acknowledge financial support from INFN, and the University of Trieste (FRA 2013). They are indebted to Prof. I. Walmsley and Prof. H. Ulbricht  for several stimulating and clarifying discussions. They also thank Dr. A. Vinante for supplying the data for the curve ``Cantilevers'' in Fig.~3, Dr. K.C. Lee for providing copy of his PhD thesis and Dr. S. Banerjee, Dr. S. Bera and Prof. T. P. Singh for a confrontation on the solutions of equation~\eqref{33}.  A final thank to Prof.  S. Modesti and Dr. D. Fausti for a clarification on phonon's dynamics.

%%%%%%%%%%%%%%%%%%%%%%%%%%%%%%%%%%%%%%%%%%%%%%%%%%%%%%%%%%%%%%%%%%%%%%%%%%%%%%%%%%%
\appendix
\section{Theoretical description of the experiment}\label{app1}
%%%%%%%%%%%%%%%%%%%%%%%%%%%%%%%%%%%%%%%%%%%%%%%%%%%%%%%%%%%%%%%%%%%%%%%%%%%%%%%%%%%
Let us first consider a single diamond, and let  ${\hat{s}}^\dagger$ and ${\hat{b}}^\dagger$ be the creation operators of a  Stokes photon and of a phonon, respectively: \begin{equation}
{\hat{s}}^\dagger |0_s, 0_b\rangle = |1_s, 0_b\rangle \quad , \quad\, {\hat{b}}^\dagger |0_s, 0_b\rangle = |0_s, 1_b\rangle \, ,
\end{equation}
where $|0_s, 0_b\rangle$ is the photonic-phononic initial vacuum state. The interaction (at time $t_s$) between the pump pulse and the diamond induces, with a small probability amplitude $\epsilon_s$, the excitation of a phononic state, accompanied by the emission of a Stokes photon. The initial vacuum state then changes to:
\begin{equation}
|\psi, t_s \rangle  = |0_s, 0_b\rangle + \epsilon_s \, |1_s, 1_b\rangle 
\end{equation}
(up to an overall normalization factor), with $|\epsilon_s|^2 \ll 1$.
For the two diamonds, we have:
\begin{eqnarray}
\label{a.a}
|\Psi, t_s \rangle &=& {|\psi, t_s \rangle}_L \otimes \, {|\psi, t_s \rangle}_R \nonumber  \\
&=& {|0_s, 0_b\rangle}_L \otimes {|0_s, 0_b\rangle}_R \, +  \epsilon_s {|1_s, 1_b\rangle}_L \otimes {|0_s, 0_b\rangle}_R + \nonumber\\
&&+ \epsilon_s  \, {|0_s, 0_b\rangle}_L \otimes {|1_s, 1_b\rangle}_R \, 
\end{eqnarray}
where the subscripts $L$ and $R$ refer to the left and right diamond respectively, and we have neglected terms proportional to $\epsilon_s^2$. Note that at this stage, there is no entanglement yet. 

Next, the Stokes modes  interfere with a relative phase shift $\phi_s$, simply resulting in an additional multiplicative factor in the right diamond's optical mode (${|1_s, 1_b\rangle}_R~\rightarrow~e^{- \, i \, \phi_s}{|1_s, 1_b\rangle}_R$), due to the different optical path in the two branches. The resulting state is still factorized.
% since we multiply by a phase factor the whole right wave function. 

The coherence lifetime of phonons is about $7$ps, but before that time, about $T = 350$fs after the pump pulse, a strong probe pulse is directed to the crystals. As a result, the phonon is converted into a $40$THz blue-shifted anti-Stokes photon with a probability $|\epsilon_a|^2 \ll 1$: $|0_a, 1_b\rangle \rightarrow |0_a, 1_b\rangle + \epsilon_a \, |1_a, 0_b\rangle $, while the vacuum state is left unaltered. Note that, in order to simplify the notation, we omitted to write explicitly the state of the Stokes photon, being irrelevant for future calculations. From now on we will use this notation.

Therefore, after the interaction with the probe pulse, the state~\eqref{a.a} changes to:
\begin{eqnarray}
\label{a.b}
\begin{aligned}
|\Psi, t_s + T\rangle = &{|0_a, 0_b\rangle}_L \otimes {|0_a, 0_b\rangle}_R  +\\
&+\epsilon_s [\,{|0_a, 1_b\rangle}_L \otimes {|0_a, 0_b\rangle}_R \, +\\
&+ e^{-i\phi_s} {|0_a, 0_b\rangle}_L \otimes {|0_a, 1_b\rangle}_R \, + \\ 
&+\, \epsilon_a  ({|1_a, 0_b\rangle}_L \otimes {|0_a, 0_b\rangle}_R +\\ 
&+ \, e^{-i\phi_s} {|0_a, 0_b\rangle}_L \otimes {|1_a, 0_b\rangle}_R )],
%|\Psi, t_s + T\rangle &= {|0_s,0_a, 0_b\rangle}_L \otimes {|0_s, 0_a, 0_b\rangle}_R  +\\
%&+\epsilon_s [\,{|1_s,0_a, 1_b\rangle}_L \otimes {|0_s,0_a, 0_b\rangle}_R \, +
%\\&+ e^{-i\phi_s} {|0_s,0_a, 0_b\rangle}_L \otimes {|1_s,0_a, 1_b\rangle}_R \, + 
%\\ 
%&+\, \epsilon_a  ({|1_s,1_a, 0_b\rangle}_L \otimes {|0_s,0_a, 0_b\rangle}_R \\ 
%&+ \, e^{-i\phi_s} {|0_s,0_a, 0_b\rangle}_L \otimes {|1_s,1_a, 0_b\rangle}_R )],
\end{aligned}
\end{eqnarray}
where, as before, higher order terms in $\epsilon_a$ have been neglected. This state is still factorized.

The next step is the creation of an entangled state. At time $t'_s > t_s + T$ the presence of Stokes photons is detected, which is a signal of a phonon being created in the crystal. The state~\eqref{a.b} is then projected to:
\begin{eqnarray}
\begin{aligned}
|\Psi, t'_s\rangle =& \epsilon_s \{  {|0_a, 1_b\rangle}_L \otimes {|0_a, 0_b\rangle}_R +\\ 
&+ \, e^{-i\phi_s} {|0_a, 0_b\rangle}_L \otimes {|0_a, 1_b\rangle}_R +\\
&+ \epsilon_a ({|1_a, 0_b\rangle}_L \otimes {|0_a, 0_b\rangle}_R +\\
&+ e^{-i\phi_s}{|0_a, 0_b\rangle}_L \otimes {|1_a, 0_b\rangle}_R ) \},
\end{aligned}
\end{eqnarray}
apart from an overall normalization factor. This state is not factorizable anymore: entanglement between the two diamonds has been created. 

%In practice, we can infer the presence of entanglement between the two diamonds due to single phonon excitation distributed across the two crystals.  
%which, neglecting the normalization factor, represents an entangled state of the two diamonds at time $t_s$ containing a single phonon excitation distribuited across the two crystals.

%When the diamonds interact with the pump pulse, an entangled state between phonons and Stokes photons is created and the latter are observed at the detector $D_s$, from which we can infer the presence of entanglement between the two diamonds, due to single phonon excitation distribuited across the two crystals.

The final step is entanglement detection. The anti-Stokes photons coming from left and right (which have different polarizations) are recombined in the same spatial mode with a polarizing beamsplitter, and interference is create, with a controllable phase $\phi_a$, by means of a half-wave plate at $45^\circ$, which deletes the polarization differences between the R and L anti-Stokes photons. This is done in order to erase the ``which way information" brought in by the polarization. After that, the flux of anti-Stokes photons is divided with another beamsplitter, which directs the incoming signal to two different detectors $D_+$ and $D_-$. The overall measurement process can be described as the projection of the out-coming state $|\Psi, t'_s\rangle$ of the states: 
\eq
| a_{\pm} \rangle \, = \, \frac{| 1_a\rangle_L \otimes | 0_a\rangle_R \, \pm\, e^{i \phi_a}\,| 0_a\rangle_L \otimes | 1_a\rangle_R  }{\sqrt 2}.
\feq

The number $ N_{\pm}$ of the anti-stokes photons counted in the two detectors is thus given by the product between the total number of incoming pulses and the measurement probability itself, which is given by:
\eq
\begin{aligned} \label{eq:fh}
P_{\pm} &= \epsilon_a^2 \left |  \langle a_{\pm}| \Psi, t'_s \rangle \right | ^2 \\
&= 2 \,  \epsilon_a^2 \, \sin^2  \left(\frac{\phi_a + \phi_s}{2} + \frac{ \pi \pm \pi }{4} \right)
\end{aligned},
\feq
where $\epsilon_a^2$ comes about because the measured quantity is not just the photo-counting in detectors $D_{\pm}$, but the coincident counts between the stokes photons detected in $D_s$ and the anti-stokes ones, which make sure that an anti-stokes photon is coming from an excited phonon state created previously with the emission of the stokes photon.
Eq.~\eqref{eq:fh} is slightly different from the one reported in \cite{lee} and corrects it, as confirmed by the authors \cite{prive}. 

The experiment confirms the theoretical prediction of Eq.~\eqref{eq:fh}, as shown in Fig.~2 of~\cite{lee} . 
As a proof of the fact that this is a measure of entanglement being created among the two diamonds, one can check what happens if the wave function of the two diamonds' phononic states system is factorized in the Left and Right branch:\- $
\left| \Psi_a \right\rangle~=~\left| \psi_L \right\rangle \,  \left| \psi_R \right\rangle$. In this case one can easily check that the count rates becomes: $ N_\pm' = \, \frac{1}{2}  \,  \epsilon_a^2 $,
meaning that all anti-stokes photons are equally divided in the two branches $+$ and $-$.

%%%%%%%%%%%%%%%%%%%%%%%%%%%%%%%%%%%%%%%%%%%%%%%%%%%%%%%%%%%%%%%%%%%%%%%%%%%%%%%%%%%
\section{Calculational details} \label{app2}
%%%%%%%%%%%%%%%%%%%%%%%%%%%%%%%%%%%%%%%%%%%%%%%%%%%%%%%%%%%%%%%%%%%%%%%%%%%%%%%%%%%
We report the calculation of the integral in Eq.~\eqref{Sandrino},  necessary to derive Eq.~\eqref{eta}, in the case  $i=j=3$. We start with the change of variables $\textbf u = \textbf x - \textbf r_1$ and  $\textbf y =~\textbf r_1 -~\textbf r_2$, obtaining:
\eq
 I_{3,3}(\textbf y) = \frac{ e^{ -\frac{\textbf y^2}{2 r_C^2} }}{r_C^4} \int d  \textbf u \  u_3 ( u_3 +y_3) e^{ -\frac{1}{r_C^2} \left( \textbf{u}^2 + \textbf u \cdot  \textbf y \ \right) } \nonumber
\feq
We decompose the integral in $d \textbf u$ in the three cartesian components $(u_1 , u_2 , u_3)$, giving  the two integrals:
\eq 
\int d u_i \,e^{-\frac{u_i}{r_C^2}\left(u_i + y_i\right)}\, = \,\sqrt{\pi} \,r_C \,e^{\frac{y_i^2}{4 r_C^2}}\,, \qquad i=1,2
\nonumber
\feq
\eq 
\int d u_3 \,e^{-\frac{u_3}{r_C^2}\left(u_3 + y_3\right)}u_3\left(u_3  + y_3 \right) = \frac{1}{4}\sqrt{\pi} r_C e^{\frac{y_3^2}{4 r_C^2}} \left(2r_C^2-y_3^2\right).
\nonumber
\feq

Finally we get the result:
\eq 
I_{3,3}\left({\textbf r}_1,{\textbf r}_2\right)= \frac{\pi^{\frac{3}{2}}}{2 r_C} e^{- \frac{\left(\textbf{r}_1-\textbf{r}_2\right)^2}{4 r_C^2}} \left[r_C^2 - \frac{\left(\textbf{r}_1-\textbf{r}_2\right)_3^2}{2}\right], 
\nonumber
\feq
which directly leads to Eq. \eqref{eta}.


\begin{thebibliography}{999}

\bibitem{aa} S. Eibenberger, S. Gerlich, M. Arndt, M. Mayor, and J. Tuxen, Phys. Chem. Chem. Phys. \textbf{15}, 14696 (2013).

\bibitem{romero} O. Romero-Isart, A. C. Pflanzer, F. Blaser, R. Kaltenbaek, N. Kiesel, M. Aspelmeyer, and J. I. Cirac. Phys. Rev. Lett., \textbf{107} (2011).

\bibitem{lee} K. C. Lee, M. R. Sprague, B. J. Sussman, J. Nunn, N. K. Langford, X. M. Jin, T. Champion, P. Michelberger, K. F. Reim, D. England, D. Jaksch and I. A. Walmsley, {Science}. \textbf{334}, 1253 (2011).

%\bibitem{agg} R.L. Aggarwal, L.W. Farrar, S.K. Saikin, X. Andrade, A. Aspuru-Guzik, D.L. Polla, {Solid State Comunication}, \textbf{152}, 3 (2012).

\bibitem{grw} G.C. Ghirardi, A. Rimini and T. Weber, Phys. Rev. D \textbf{34}, 470 (1986).

\bibitem{pearle} P. Pearle, Phys. Rev. A \textbf{39}, 2277 (1989). 

%\bibitem{bass1} A. Bassi, E. Ippoliti and  S.L. Adler, J. Phys. A: Math. Gen. \textbf{38}, 2715 (2005).

\bibitem{gpr} G.C. Ghirardi, P. Pearle and A. Rimini, Phys. Rev. A \textbf{42}, 78 (1990).

\bibitem{diosi} L. Diosi, Phys. Rev. A \textbf{40}, 1165 (1989). 

\bibitem{gatarek} D. Gatarek and N. Gisin, J. Math. Phys. \textbf{32}, 2152 (1991). 

\bibitem{chru} D. Chruscinski and P. Staszewski, Physica Scripta \textbf{45}, 193 (1992). 

\bibitem{bass2} A. Bassi and G. C. Ghirardi, Phys. Reports \textbf{379}, 257 (2003).

\bibitem{bass3} A. Bassi, E. Ippoliti, and S. L. Adler, Phys. Rev. Lett. \textbf{94}, 030401 (2005).

\bibitem{bass4} A. Bassi, K. Lochan, S. Satin, T. P. Singh, and H. Ulbricht, Rev. Mod. Phys. \textbf{85}, 471 (2013).

\bibitem{adler} S. L. Adler, J. Phys. A, \textbf{40}, 13501 (2007).

\bibitem{jo-ze} E. Joos and H.D. Zeh, Z. Phys. B \textbf{59}, 223 (1985).

\bibitem{prive} I. Walmsley, private communication.

\bibitem{nhh} S. Nimmrichter, K. Hornberger and K. Hammerer, Phys. Rev. Lett. \textbf{113}, 020405 (2014).


\bibitem{cc} C. Curceanu, B.C. Hiesmayr, and K. Piscicchia, J. Adv. Phys. \textbf{4}, 263 (2015).

\bibitem{bb} A. Vinante, M. Bahrami, A. Bassi, O. Usenko, G. Wijts and T.H. Oosterkamp, \textit{Upper bounds on spontaneous wave-function collapse models using millikelvin-cooled nanocantilevers}, arXiv:1510.05791 (2015).

\bibitem{mt} M Toros and A. Bassi, {\it Bounds on Collapse Models from Matter-Wave Interferometry},  arXiv:1601.03672.
 
\end{thebibliography}
\end{document}